\documentclass[a4paper,11pt]{article}
\usepackage{pos}

\title{Flavor and Fluctuations}

\author*[a]{Stam Nicolis}

\affiliation[a]{Institut Denis Poisson, Université de Tours, Université d'Orléans, CNRS (UMR7013)\\
 Parc Grandmont, Tours 37200, France}

\emailAdd{stam.nicolis@lmpt.univ-tours.fr}

\abstract{The Parisi-Sourlas approach to supersymmetry implies that, in spacetime dimensions greater than 1, there is a constraint on the minimal number of flavors, in order for a field theory to define a closed system. In particular, this number is greater than 1.  This does not preclude that supersymmetry can be broken, however, and the known ways of breaking supersymmetry can be taken into account  from this point of view, by using the so-called Nicolai map. This procedure is well-defined for abelian gauge theories and corresponds to the construction of the so-called trivializing maps for non-abelian gauge theories, that is, still, work in progress.  }

\FullConference{Corfu Summer Institute 2023 "School and Workshops on Elementary Particle Physics and Gravity" (CORFU2023)\\
 23 April - 6 May , and 27 August - 1 October, 2023\\
Corfu, Greece\\}

\begin{document}
\maketitle

\section{Introduction}\label{intro}
The notion of ``flavor'' entered physics with the discovery of the muon in 1935; and Rabi's question, ``Who ordered that?'' has remained unanswered since. This hasn't changed with the discovery of more flavors, nor of the intricate relations between the flavors. 

There are two issues to the flavor problem: (a) What are the constraints on the number of flavors and (b) what are the constraints on the masses of the flavors, what ``stabilizes'' the flavor hierarchy. 

Of course at the level of the classical action, the number of flavors isn't constrained in any way, it can take any integer value. 

It's when quantum effects must be taken into account, that constraints on the number of flavors may appear, though, even then, these constraints don't imply any resolution of the problem of the flavor hierarchy itself: QED is a consistent quantum field theory of any number of electrically charged fermions, interacting with photons and doesn't seem  to require the existence of any particular number of  flavors for its consistency (at least in perturbation theory; it isn't known what happens beyond perturbation theory). If any number of flavors is  included, no particular constraints on the masses of the flavors, in relation to each other seem to emerge, either. This persists in the Standard Model, i.e. when we include weak and strong interactions, also, since there doesn't appear to be any fundamental reason for more than one family: The number of families (and the mass hierarchies) seems to be determined by experiment, not theory. 

However the stochastic approach to the description of quantum fluctuations, sketched by Parisi and Sourlas in ref.~\cite{parisi_sourlas} seems to offer a new way of framing the problem and seems to provide a reason for the existence of more than one families, when the worldvolume is assumed to have certain symmetries, so, in the present contribution, we shall try to study some of its consequences, that are of relevance for understanding how a flavor hierarchy can arise. The key ingredient is the so-called ``Nicolai map''~\cite{Nicolai:1980jc,Nicolai:1980js}. 

The example we shall use to illustrate the idea will be the $\mathcal{N}=2,$ $D=2$ Wess-Zumino model. In the next section we shall recall its definition, in the particular formalism of Parisi and Sourlas and in the section following we shall describe the generalization, that can describe a flavor hierarchy. The challenges involved in generalizing the approach to more than two spacetime dimensions will be outlined in the conclusions. 

\section{The $\mathcal{N}=2,$ $D=2$ Wess-Zumino model à la Parisi-Sourlas}\label{WZ-PS}
In ref.~\cite{parisi_sourlas} Parisi and Sourlas remarked that it is possible to describe the $\mathcal{N}=2,D=2$ Wess--Zumino model in the following way: Define two scalar fields, $\eta_1$ and $\eta_2$ by the expressions
\begin{equation}
\label{noisefields}
\begin{array}{l}
\displaystyle
\eta_1 = \partial_x\phi_2 + \partial_y\phi_1 + g(\phi_1^2-\phi_2^2)\\
\displaystyle
\eta_2 = \partial_x\phi_1 - \partial_y\phi_2 + 2g\phi_1\phi_2
\end{array}
\end{equation}
where the $\phi_1$ and $\phi_2$ are, also, scalar fields. Here $g$ is a coupling constant.

Now we construct the combination
\begin{equation}
\label{eta1eta2so2}
\begin{array}{l}
\displaystyle
\frac{1}{2}\left( \eta_1^2+\eta_2^2\right)  = \frac{1}{2}\left(
(\partial_x\phi_1)^2+(\partial_y\phi_1)^2+(\partial_x\phi_2)^2+(\partial_y\phi_2)^2
\right)+
\frac{g^2}{2}\left(\phi_1^2+\phi_2^2\right)^2 + \\
\displaystyle
\hskip2.5truecm
\partial_x\phi_2\partial_y\phi_1 - \partial_x\phi_1 \partial_y\phi_2 + g(\phi_1^2-\phi_2^2)(\partial_x\phi_2 + \partial_y\phi_1)+2g\phi_1\phi_2(\partial_x\phi_1 - \partial_y\phi_2)
\end{array}
\end{equation}
and we remark that it consists of three parts, two of which are invariant under SO(2) coordinate transformations and SO(2) ``flavor'' transformations--that act on the indices of the scalars, $\phi_I;$ and a third term that is a total derivative 
\begin{equation}
\label{totder}
\begin{array}{l}
\displaystyle
\partial_x\phi_2\partial_y\phi_1 - \partial_x\phi_1 \partial_y\phi_2 + g(\phi_1^2-\phi_2^2)(\partial_x\phi_2 + \partial_y\phi_1)+2g\phi_1\phi_2(\partial_x\phi_1 - \partial_y\phi_2)=\\
\displaystyle
\partial_x\left(\phi_2\partial_y\phi_1\right)-\partial_y\left(\phi_2\partial_x\phi_1\right) + g\left( 
\partial_x\left(\phi_1^2\phi_2-\frac{\phi_2^3}{3} \right) +\partial_y\left( \frac{\phi_1^3}{3}-\phi_1\phi_2^2 \right)
 \right)
\end{array}
\end{equation}
and, therefore, may be dropped, upon imposing periodic boundary conditions. 

We now {\em assume} that the partition function for the fields, $\eta_1$ and $\eta_2$ is given by the expression
\begin{equation}
\label{Zeta}
Z = \int\,[\mathscr{D}\eta_1][\mathscr{D}\eta_2]\,e^{-\int\,d^2x\,\frac{1}{2}\left( \eta_1^2+\eta_2^2\right)}
\end{equation}
which can be taken equal to 1 by a convenient definition of the integration measure. 

What Parisi and Sourlas pointed out is that, if we interpret eqs.~(\ref{noisefields}) as an injunction to change variables in the expression~(\ref{Zeta}), in order to obtain the partition function for the $\phi_1,\phi_2,$ we obtain the expression
\begin{equation}
\label{Zphi}
\begin{array}{l}
\displaystyle
Z = \int\,[\mathscr{D}\phi_1][\mathscr{D}\phi_2]\,\left|\mathrm{det}\,\frac{\delta\eta_I}{\delta\phi_J}\right|\,e^{-S[\phi_1,\phi_2]}=1
\end{array}
\end{equation}
The classical action of the scalars, $S[\phi_1,\phi_2],$ is given by the expression
\begin{equation}
\label{Sscalar}
S[\phi_1,\phi_2]=\frac{1}{2}\left(
(\partial_x\phi_1)^2+(\partial_y\phi_1)^2+(\partial_x\phi_2)^2+(\partial_y\phi_2)^2
\right)+
\frac{g^2}{2}\left(\phi_1^2+\phi_2^2\right)^2
\end{equation}
The Jacobian is given by the expression
\begin{equation}
\label{Jacobian}
\left|\mathrm{det}\,\frac{\delta\eta_I}{\delta\phi_J}\right|=e^{-\mathrm{i}\theta_\mathrm{det}}\,\mathrm{det}
\left(\begin{array}{cc} \partial_y+2g\phi_1 & \partial_x - 2g\phi_2\\ \partial_x+2g\phi_2 & -\partial_y+2g\phi_1\end{array}  \right)
\end{equation}
We remark that the differential operator, whose determinant appears here, is a local operator.
Therefore, upon introducing the determinant of the differential operator in the action, using anticommuting fields, we realize that these fields 
describe (a) target space fermions and (b) these fermions are the superpartners of the scalars, in a way that's consistent with the $\mathcal{N}=2$ suprersymmetry algebra~\cite{parisi_sourlas}. The term proportional to $g$ describes the interactions between the fermions and the scalars. 

The ``bonus'' is the appearance of the phase of the determinant, $e^{-\mathrm{i}\theta_\mathrm{det}},$ which implies that the partition function of the fields $\eta_1,\eta_2,$ upon perfoming the change of variables~(\ref{noisefields}), becomes the Witten index of the $\mathcal{N}=2,D=2$ Wess--Zumino model; and the relations~(\ref{noisefields}) can be understood as defining the Nicolai map for the $\mathcal{N}=2,D=2$ Wess--Zumino model. 

So imposing, in fact, that the Jacobian should be a Dirac operator with a particular interaction, inevitably leads to the identification of the noise fields and the emergence of a particular class of supersymmetric theory. Indeed the reasoning by Nicolai that led to the ``Nicolai map'' is precisely this, in reverse: One starts from a supersymmetric theory, integrates out the fermions and identifies the determinant with the Jacobian of a change of variables in the path integral. What Parisi and Sourlas realized is that the argument can be run in the other direction, also, and that one can obtain a supersymmetric theory by expressing the fluctuations of fields by their superpartners, within a class of supersymmetric theories, namely those with extended supersymmetry. More precisely, the fluctuations of the fields are ``resolved'' by their superpartners, within a certain class of supersymmetric theories, namely those with $\mathcal{N}=2$ supersymmetry.

In particular the number of fields and their masses are fixed. The non-trivial statement is that we had to have, in the example at hand, at least two scalars and thus two flavors. The hierarchy is very simple, since supersymmetry imposes that the masses vanish. In the next section we shall present the generalization to massive fields and a less trivial hierarchy.

\section{Inroducing a hierarchy}\label{hierflav}
We remark that eqs.~(\ref{noisefields}) lead to a scalar potential,
\begin{equation}
\label{Vscal}
V(\phi_1,\phi_2) = \frac{g^2}{2}(\phi_1^2+\phi_2^2)^2
\end{equation}
that has its minimum at the origin. So the question arises, how might it be possible to break the SO(2) flavor symmetry. 

One way is by realizing that, when writing the equations for the $\eta_1,\eta_2$ in terms of $\phi_1$ and $\phi_2$ we weren't as  general as we could be, if we impose, just, invariance under SO(2) coordinate transformations. We can add linear terms, {\em i.e.} write the following expressions for the Nicolai map:
\begin{equation}
\label{noisefields1}
\begin{array}{l}
\displaystyle 
\eta_1 = \partial_x\phi_2 + \partial_y\phi_1 + g(\phi_1^2-\phi_2^2)+c_1\phi_1+c_2\phi_2\\
\displaystyle
\eta_2 = \partial_x\phi_1 - \partial_y\phi_2 + 2g\phi_1\phi_2+d_1\phi_1+d_2\phi_2
\end{array}
\end{equation}
The constraints on the coefficients $c_1,c_2$ and $d_1,d_2$ are deduced from the requirement that the cross-terms  (that aren't invariant under SO(2) coordinate transformations) are total derivatives.  A straightforward calculation leads to the conclusion that the coefficients of the linear terms must satisfy the relations
\begin{equation}
\label{linearterms}
c_1=d_2\equiv C\,\hskip0.5truecm \mathrm{and}\,\hskip0.5truecm c_2=-d_1\equiv -D
\end{equation}
The scalar potential now is given by the expression
\begin{equation}
\label{Vscalar1}
V(\phi_1,\phi_2) = \frac{g^2}{2}(\phi_1^2+\phi_2^2)^2+\frac{(C^2+D^2)}{2}(\phi_1^2+\phi_2^2) + g(\phi_1^2-\phi_2^2)(C\phi_1-D\phi_2) + 2g\phi_1\phi_2(D\phi_1+C\phi_2)
\end{equation}
We remark that we may set $C\equiv\kappa\cos\alpha$ and $D\equiv\kappa\sin\alpha$ and, upon adopting a polar representation for the scalars, $\phi_1\equiv\phi\cos\theta$ and $\phi_2\equiv \phi\sin\theta,$ where $\theta$ is a putative mixing angle, the scalar potential takes the form
\begin{equation}
\label{Vscalarpolar}
V=\frac{\kappa^2}{2}\phi^2+\frac{g^2}{2}\phi^4+g\kappa\phi^3\cos(\theta-\alpha)
\end{equation}
we realize that it is possible to interpolate between different phases, where supersymmetry and flavor symmetry are both realized, both broken, or one of them is realized and the other is broken. This is, of course, in line with what was understood fifty years ago~\cite{Fayet:1975ki,ORaifeartaigh:1975nky}; what is new is that that analysis (as, also,  the introduction of the Nicolai map~\cite{Nicolai:1980jc,Nicolai:1980js}) assumed supersymmetry as an imput; whereas the idea of Parisi and Sourlas--as pursued here--is, rather, that supersymmetry is an output--the new input is that fluctuations are resolved by dynamical degrees of freedom, that can be identified as the superpartners of the ``dynamical'' degrees of freedom that appear in the classical action. This is at least one way that supersymmetry can be ``hidden'' in a non-trivial way; and can be ``revealed'' by computing the correlation functions of the ``noise fields'', $\eta_1$ and $\eta_2,$ expressed in terms of the scalars (if the dynamical degrees of freedom are assumed to be scalars), without the need of referring to the fermions explicitly at all--which was the idea put forward by Nicolai, for supersymmetric theories, but, now, understood as applying to non-supersymmetric theories, as well. That this is possible was checked, by preliminary numerical simulations  in ref.~\cite{Nicolis:2017lqk}.  So it isn't necessary to start with the $\mathcal{N}=2,D=2$ Wess-Zumino model; it suffices to start with a theory of two scalar fields and the fermions will emerge as resolving the fluctuations.  Including the terms, proportional to $\kappa,$ may break the flavor symmetry, supersymmetry, both or neither and, thus, lead to a hierarchy, whose stability is controlled by the ``hidden'' supersymmetry. What is of interest is that, if one does start with only one (i.e. real)  scalar field, one finds that it isn't possible to write a Nicolai map of the form~(\ref{noisefields}) or~(\ref{gnnoisefields})--the ``completion'' is ambiguous and it doesn't seem possible to identify the superpartners as resolving the fluctuations. 
\section{Conclusions}\label{concl}
Our analysis has been carried out within the framework of the $\mathcal{N}=2,D=2$ Wess--Zumino model;  however its scope is much broader. While in the traditional approach towards supersymmetry one starts with this model and explores the various ways supersymmetry and/or the flavor symmetry can be realized or broken~\cite{Fayet:1975ki,ORaifeartaigh:1975nky}, in the approach proposed by Parisi and Sourlas, supersymmetry is, in fact, an output of the analysis, rather than an input.  

The input is the Nicolai map~(\ref{noisefields}) and its generalization~(\ref{noisefields1}). And its structure ``conceals'', both  the spacetime symmetry--in the present case invariance under global SO(2) coordinate transformations--{\em and}  the flavor symmetry, along with the putative flavor hierarchy.   In so doing, it ``conceals'' the fact that, in two spacetime dimensions, at least, two scalars are required--which, in turn, implies that four fermions are required and, thus, at least, two flavors: The reason the derivative terms in eq.~(\ref{noisefields}),(\ref{noisefields1}) appear as they do is due to the fact that the Jacobian of the transformation betwen the $\eta_1$ and $\eta_2$ and the $\phi_1$ and $\phi_2$ must be a Dirac operator, in order that target space fermions appear--and, thus, target space supersymmetry.   This shows one way,  how flavor might emerge. In previous work~\cite{Nicolis:2017lqk,Nicolis:2023mre} and in the present contribution, we have tried to explore the consequences of the idea of Parisi and Sourlas and of Nicolai, in an effort to better show how supersymmetry can be ``hidden in plain sight''. There is much more work to be done to clarify its properties. It should be stressed that the Nicolai map isn't a ``Langevin equation'', though it may have been called that in the literature: The difference is that the Nicolai map describes fluctuations {\em at equilibrium}, while the Langevin equation describes fluctuations {\em towards equilibrium}\footnote{I'm grateful to A. Schwimmer for discussions on this point.}

Of course an immediate question is, how might these considerations be generalized to other spacetime dimensions, in particular the four spacetime dimensions of the real world. The obstruction identified by Parisi and Sourlas~\cite{parisi_sourlas} is due to the fact that, in Euclidian signature, if $D\equiv\hskip-0.29truecm/\,2\,\mathrm{mod}\,8,$ the Dirac matrices do not have a Majorana representation and the solution is to double the degrees of freedom, as sketched in~\cite{Nicolis:2023mre,Nicolis:2021buh}. This is necessary, however does not preclude the appearance of further obstacles, namely that the cross products, that were total derivatives in two dimensions, are, also, total derivatives in four dimensions. Such an obstruction seems to appear, already in the absence of any superpotential, so cannot be blamed on it. There's something here that isn't, yet,  fully understood. 

Furthermore, while taking into account abelian gauge fields doesn't seem to present issues of principle (cf.~\cite{Nicolis:2023mre} for a way to do this), non-abelian gauge fields present a challenge, still, since the Nicolai map isn't known (attempts to construct it, under the name of ``trivializing map'', started,after first attempts in the 1980s~\cite{deAlfaro:1982ex,DeAlfaro:1986uv}, with the work of Lüscher~\cite{luscher2010trivializing,luscher2015instantaneous} and have been pursued  in refs.~\cite{Lechtenfeld:2021yjb,Lechtenfeld:2021uvs,Ananth:2020gkt,Ananth:2020lup}, without, however, as conclusive a result as for Wess--Zumino models in two dimensions. 
  
{\bf Acknowledgements:} It's a pleasure to thank M. Axenides, P. Fayet, E. Floratos, J. Iliopoulos and A. Schwimmer for their interest and for stimulating discussions. I would also like to thank the organizers of the 2023 Corfu Workshops and the organizers of the Rencontres de Physique des Particules 2024 for the opportunity of presenting these ideas and the participants for their questions.  This research is part of the research project ``Chaotic behavior of closed quantum systems'', supported by the CNRS International Emerging Actions program, under contract 318687. 
\bibliographystyle{JHEP}
\bibliography{SUSY}

\end{document}